\documentclass[a4paper,11pt]{article}

\pdfoutput=1
\usepackage{jcappub}

\usepackage{bbm}
\usepackage{mathrsfs}
\usepackage{slashed}
\usepackage{caption}
\usepackage{epstopdf}
\usepackage[normalem]{ulem}
\usepackage[bottom]{footmisc}
\usepackage{subcaption}
\usepackage{bbold}
\usepackage{titlesec}
\usepackage{threeparttable}
\usepackage{booktabs}
\usepackage{changepage}
\usepackage[utf8]{inputenc}
\usepackage{dsfont} 
\usepackage{grffile}
\usepackage{graphicx}
\usepackage{dcolumn}
\usepackage{bm}
\usepackage{amssymb}
\usepackage{setspace}
\usepackage{amsmath}
\usepackage{array}
\usepackage{float}
\usepackage{lmodern}
\usepackage{multirow}
\usepackage{soul}
\usepackage{braket}
\usepackage{comment}
\usepackage[draft]{pgf}
\usepackage{adjustbox} 
\usepackage{xspace} 
\usepackage{url}
\usepackage{xcolor}
\usepackage{orcidlink}
\usepackage{chngcntr}
\usepackage{indentfirst}

\allowdisplaybreaks

\newcommand{\be}{\begin{equation}}
\newcommand{\ee}{\end{equation}}

\newcommand{\bea}{\begin{eqnarray}}
\newcommand{\eea}{\end{eqnarray}}

\title{Probing Fundamental Constant Oscillation in the Galactic Center with S-Star Spectroscopy}

\author{
Zhaoyu Bai$^{a}\,\orcidlink{0000-0002-0758-1579}$,
Vitor Cardoso$^{b,c}\,\orcidlink{0000-0003-0553-0433}$,
Yifan Chen$^{d,e}$\footnote{Corresponding authors.}\,\orcidlink{0000-0002-2507-8272},
Tuan Do$^{f}$\,\orcidlink{0000-0001-9554-6062},
Aur\'elien Hees$^{g}$\footnotemark[1]\,\orcidlink{0000-0002-2186-644X},
Huangyu Xiao$^{h,i}\,\orcidlink{0000-0003-2485-5700}$,
and Xiao Xue$^{j}\,\orcidlink{0000-0002-0740-1283}$
}

\affiliation{
$^a$Department of Particle Physics and Astrophysics, Weizmann Institute of Science, Rehovot 7610001, Israel\\
$^b$Center of Gravity, Niels Bohr Institute, Blegdamsvej 17, 2100 Copenhagen, Denmark\\
$^c$CENTRA, Departamento de F\'{\i}sica, Instituto Superior T\'ecnico (IST), Universidade de Lisboa (UL), Avenida Rovisco Pais 1, 1049 Lisboa, Portugal\\
$^d$State Key Laboratory of Dark Matter Physics, Tsung-Dao Lee Institute, Shanghai Jiao Tong University, Shanghai 200240, China\\
$^e$Key Laboratory for Particle Astrophysics and Cosmology (MOE) \& Shanghai Key Laboratory for Particle Physics and Cosmology, Shanghai Jiao Tong University, Shanghai 200240, China\\
$^f$Physics and Astronomy Department, University of California, Los Angeles, 475 Portola Plaza, Los Angeles, CA 90095, USA\\
$^g$LTE, Observatoire de Paris, Universit\'e PSL, Sorbonne Universit\'e, Universit\'e de Lille, LNE, CNRS, 61 Avenue de l'Observatoire, 75014 Paris, France\\
$^h$Kavli Institute for Cosmological Physics, University of Chicago, Chicago, IL 60637, USA\\
$^i$Astrophysics Theory Department, Theory Division, Fermilab, Batavia, IL 60510, USA\\
$^j$Institut de F\'{\i}sica d’Altes Energies (IFAE), The Barcelona Institute of Science and Technology, Campus UAB, 08193 Bellaterra (Barcelona), Spain
}

\emailAdd{zhaoyu.bai@weizmann.ac.il}
\emailAdd{vitor.cardoso@nbi.ku.dk}
\emailAdd{chen.yifan@sjtu.edu.cn}
\emailAdd{tdo@astro.ucla.edu}
\emailAdd{aurelien.hees@obspm.fr}
\emailAdd{huangyu@fnal.gov}
\emailAdd{xxue@ifae.es}

\abstract{
Astrophysical spectroscopy provides a powerful probe of spacetime variations of fundamental constants, as atomic and ionic emission and absorption lines depend sensitively on the fine-structure constant. In particular, coherent temporal oscillations induced by an ultralight scalar background produce characteristic, time-resolved signatures that can be robustly disentangled from intrinsic variability. In the Galactic Center, such scalar backgrounds can be substantially enhanced, either through the formation of dense scalar clouds powered by black hole rotational energy extraction or as ultralight scalar dark matter forming a soliton-like core. These scalar configurations generically induce oscillations of the fine-structure constant, with periods set by the scalar mass and spatial profiles determined by the scalar wavefunction and its coupling to the electromagnetic sector. We show that precise, time-resolved spectroscopy of S-stars orbiting the supermassive black hole Sgr A$^*$ provides a sensitive test of these effects, enabling constraints on quadratic scalar–photon couplings in the exceptionally high boson-density environment of the Galactic Center.

\bigskip

}

\begin{document}

\maketitle
\flushbottom

\section{Introduction}

Ultralight bosons, including the QCD axion originally proposed to solve the strong CP problem~\cite{Preskill:1982cy,Abbott:1982af,Dine:1982ah} and more broadly predicted in theories with extra dimensions~\cite{Svrcek:2006yi,Arvanitaki:2009fg}, are well-motivated candidates for physics beyond the Standard Model. Spanning a wide mass range from $\sim10^{-22}$ to $1~\mathrm{eV}$, such particles can constitute dark matter (DM) and behave as coherently oscillating classical fields with large occupation numbers. This wave-like nature gives rise to distinctive dynamical and interference phenomena, motivating a broad program of terrestrial experiments and astrophysical observations to search for their signatures.

Searches for ultralight bosons often target their interaction portals with Standard Model sectors. A particularly precise approach probes modulations of the electromagnetic fine-structure constant, $\alpha_{\rm EM}$, induced by couplings to the kinetic term $F_{\mu\nu}F^{\mu\nu}$~\cite{Uzan:2002vq,Olive:2007aj,Uzan:2010pm,Stadnik:2015kia,Martins:2017yxk,Safronova:2017xyt,Uzan:2024ded}. Such modulations arise directly for dilatons~\cite{Taylor:1988nw,Damour:1994ya,Antoniadis:1998ig,Damour:2010rm,Damour:2010rp} and relaxions~\cite{Graham:2015cka,Flacke:2016szy}, or at loop level for axions~\cite{Kim:2022ype,Beadle:2023flm,Kim:2023pvt}. Terrestrial experiments, including atomic spectroscopy~\cite{Arvanitaki:2014faa,VanTilburg:2015oza,Hees:2016gop,Zhang:2022ewz} and optomechanical systems~\cite{Stadnik:2014tta,Stadnik:2015xbn,Branca:2016rez,Campbell:2020fvq,Vermeulen:2021epa}, have achieved unprecedented sensitivity to variations in $\alpha_{\rm EM}$.

Recent progress in astrophysical spectroscopy has enabled tests of the fine-structure constant at the Galactic Center~\cite{Hees:2020gda}. While current observations lack the precision of laboratory experiments, the signal can be strongly enhanced by the high density of ultralight bosons in this region. Mechanisms such as superradiance~\cite{Penrose:1971uk,ZS,Brito:2015oca} or relaxation driven by self-interactions~\cite{Budker:2023sex,Gan:2023swl} can generate dense boson clouds around compact objects, with field amplitudes approaching $\sim10^{16}$~GeV~\cite{Chen:2022kzv,Chen:2023vkq,Ayzenberg:2023hfw}. The vicinity of black holes (BHs) therefore provides a particularly powerful environment to probe ultralight bosons in the relevant mass range.

In this work, we propose using high-precision spectroscopic observations of stars orbiting the supermassive BH (SMBH) Sgr A$^*$ to search for ultralight scalar fields with quadratic couplings to the electromagnetic sector. In the case of axions, superradiant amplification around the rotating SMBH can drive the field amplitude toward saturation near the decay constant, leading to predictable, time-dependent variations of the fine-structure constant in the vicinity of Sgr A$^*$. We also consider scenarios in which ultralight scalar DM forms a soliton-like core at the Galactic Center.

\section{Quadratic Scalar Couplings to Photons}

As pseudoscalars, axions $\phi$ couple linearly to parity-odd operators such as $\phi\,G_{\mu\nu}\tilde G^{\mu\nu}$ and $\phi\,F_{\mu\nu}\tilde F^{\mu\nu}$, where $G_{\mu\nu}$ and $F_{\mu\nu}$ are the gluon and photon field strength tensors. At one loop, however, quadratic couplings to the electromagnetic kinetic term are generically induced~\cite{Kim:2022ype,Beadle:2023flm,Kim:2023pvt},
\begin{equation}
    \mathcal{L}_{\text{EM}} =  \frac{C_\gamma}{4} \frac{\phi^2}{f_\phi^2} F_{\mu\nu} F^{\mu\nu},\label{eq:aaFF}
\end{equation}
where $f_\phi$ is the axion decay constant and $C_\gamma$ is a dimensionless coefficient. For ALPs, this coupling depends on the ultraviolet realization of shift-symmetry breaking and $C_\gamma$ may be treated as a free parameter~\cite{Beadle:2023flm}. 

{A particularly well-motivated class of models consists of axions that possess the standard dimension-five coupling to gluons,
\begin{equation}
\mathcal L \supset \frac{\phi}{f_\phi}\frac{\alpha_s}{8\pi}
G_{\mu\nu}\tilde G^{\mu\nu},
\end{equation}
where $\alpha_s$ is the QCD gauge coupling.
In this case, pion–axion mixing together with hadronic loop effects generates a quadratic coupling to photons of the form in Eq.~(\ref{eq:aaFF}), yielding the robust prediction
\begin{equation}
C_\gamma^{\rm QCD}
\simeq
-3\times10^{-5},
\label{eq:CgammaQCD}
\end{equation}
with only mild sensitivity to ultraviolet details~\cite{Kim:2022ype,Beadle:2023flm,Kim:2023pvt}. Throughout this work, we use $C_\gamma^{\rm QCD}$ to denote the characteristic value generated from the axion-gluon coupling.

For the minimal QCD axion, the axion mass $\mu$ and decay constant $f_\phi$ are related through the QCD scale, $\mu f_\phi \sim (100\,{\rm MeV})^2$.
However, this relation can be modified in models with discrete $\mathbb Z_N$ shift symmetries, which allow parametrically smaller $f_\phi$ at fixed $\mu$ while preserving the axion solution to the strong CP problem~\cite{Hook:2018jle}. Motivated by these constructions, we treat $\mu$ and $f_\phi$ as independent parameters and consider a broader ultralight parameter space than that of the minimal QCD axion.

More generally, similar quadratic couplings to photons can arise for scalar fields protected by symmetries such as $\mathbb{Z}_2$, parameterized as \be \mathcal{L} \supset \frac{1}{4\Lambda^2} \phi^2 {F_{\mu\nu} F^{\mu\nu}},\ee with $\Lambda$ the ultraviolet scale~\cite{Stadnik:2014tta,Stadnik:2015kia,Stadnik:2015xbn,Stadnik:2016zkf,Hees:2018fpg,Hees:2019nhn,Masia-Roig:2022net,Bouley:2022eer,Banerjee:2022sqg,Gan:2023wnp,Brax:2023udt}.} Identifying $C_\gamma/f_\phi^2 \leftrightarrow 1/\Lambda^2$, we express both cases in a common framework. We then focus on quadratic couplings, since linear scalar interactions are already strongly constrained by fifth-force and equivalence-principle tests~\cite{Berge:2017ovy,Hees:2018fpg}.
Quadratic interactions evade these bounds by inducing fifth forces only at loop level. If the scalar does not make up most of the DM, the dominant constraint arises from SN\,1987A cooling, requiring $\Lambda=f_\phi/|C_\gamma|^{1/2}\gtrsim 3$~TeV~\cite{Olive:2007aj,Cox:2024oeb,Gan:2025nlu}.

Relative to the canonical electromagnetic term, $\mathcal{L}\supset -F_{\mu\nu}F^{\mu\nu}/4$, Eq.~(\ref{eq:aaFF}) and related scalar couplings induce shifts in the fine-structure constant,
\begin{equation}
    \frac{\delta \alpha_{\text{EM}}}{\alpha_{\text{EM}}} \simeq C_\gamma\frac{\phi^2}{f_\phi^2} \quad \text{or} \quad \frac{\phi^2}{\Lambda^2},\label{eq:deltaalpha}
\end{equation}
assuming $|\delta\alpha_{\rm EM}|\ll\alpha_{\rm EM}$.

\section{Fine-Structure Constant Oscillations at the Galactic Center}
\subsection{Superradiant Scalar Cloud}

Ultralight scalars with a Compton wavelength exceeding the BH horizon can form quasi-bound clouds around Kerr BHs. These hydrogen-like states are labeled by quantum numbers $(n, \ell, m)$, representing the principal, orbital angular, and azimuthal numbers, respectively~\cite{Detweiler:1980uk}. The size of the cloud is governed by the gravitational fine-structure constant $\alpha_G \equiv G_{\rm N} M_{\rm BH} \mu$, where $G_{\rm N}$ is Newton’s constant and $M_{\rm BH}$ is the BH mass. The rapid rotation of BHs can select specific bound states, transitioning them from decaying states due to the horizon to exponentially growing modes~\cite{Penrose:1971uk,ZS,Detweiler:1980uk,Cardoso:2005vk,Dolan:2007mj,Brito:2015oca}. 

This phenomenon, known as superradiance, occurs when the horizon rotates faster than the angular phase velocity of the bound state. This condition can be written as
\begin{equation}
\frac{\alpha_G}{m} < \frac{a_J}{2\left(1+\sqrt{1-a_J^2}\right)},
\label{eq:SRC}
\end{equation}
where $a_J\leq1$ is the dimensionless BH spin. For minimally coupled scalars, superradiant growth continues until it significantly spins down the BH, producing a cloud mass $M_{\rm cloud}$ up to $\sim10\%$ of $M_{\rm BH}$~\cite{Brito:2014wla,East:2017ovw,Herdeiro:2021znw} and field amplitudes approaching $\sim10^{16}$~GeV, making BH environments sensitive probes of ultraviolet physics~\cite{Chen:2022kzv,Chen:2023vkq,Ayzenberg:2023hfw,Lyu:2025lue}.

When the axion decay constant lies below $\sim10^{16}$~GeV, strong self-interactions from the cosine potential $V=-\mu^2f_\phi^2\cos(\phi/f_\phi)$ terminate the exponential growth before significant BH spin-down~\cite{Arvanitaki:2010sy,Yoshino:2012kn,Gruzinov:2016hcq,Fukuda:2019ewf,Baryakhtar:2020gao,Omiya:2020vji,Omiya:2022mwv,Omiya:2022gwu,Collaviti:2024mvh,Takahashi:2024fyq,Aurrekoetxea:2024cqd,Witte:2024drg,Miller:2025yyx,Xie:2025npy}. In the Newtonian limit $\alpha_G\ll1$, self-interactions drive a saturated state of the $(2,1,1)$ and $(3,2,2)$ modes, where weak axion leakage from $(3,2,2)$ annihilation balances superradiant injection~\cite{Gruzinov:2016hcq,Baryakhtar:2020gao}. 

{Since the $(2,1,1)$ mode grows fastest and dominates the saturated field, we focus on this mode, whose wavefunction is~\cite{Gruzinov:2016hcq,Baryakhtar:2020gao}
\be \phi = \phi_0^{\rm max}\, R(r) \cos \left(\mu t - \varphi+\Delta_0\right) \sin \theta,\label{eq:phiwave}\ee
where $(t,r,\theta,\varphi)$ are spherical coordinates with $\theta=0$ aligned with the BH spin axis and $\Delta_0$ is an arbitrary initial phase. The normalized radial wavefunction is
\begin{equation}
R(r)
=
e^{1-\frac{\alpha_G^2 r}{2r_g}}
\frac{\alpha_G^2 r}{2r_g},
\end{equation}
which ranges from $0$ to $1$. Here $r_g\equiv G_NM_{\rm BH}$ is the gravitational radius. The maximum field amplitude occurs at $r=2r_g/\alpha_G^2$ and $\theta=\pi/2$, and is related to the cloud mass through
\begin{equation}
\phi_0^{\rm max}
\simeq
0.07\,\alpha_G^2 M_{\rm pl}
\sqrt{\frac{M_{\rm cloud}}{M_{\rm BH}}},
\end{equation}
obtained by integrating the cloud energy density in the Newtonian limit, where
$M_{\rm pl}\equiv G_N^{-1/2}$ is the Planck mass.

For a self-interaction-saturated cloud, the maximum amplitude approaches
\begin{equation}
\phi_0^{\rm max}
\simeq
\frac{1}{2}\,\alpha_G f_\phi.
\label{eq:phimax}
\end{equation}
Within the superradiant mass window, the BH spin and superradiant growth rate have little influence on the Newtonian wavefunction and saturated field amplitude, as the latter is primarily determined by self-interaction-driven multi-mode dynamics~\cite{Gruzinov:2016hcq,Baryakhtar:2020gao}.}

\begin{figure}[h]
\centering
\includegraphics[width=0.75\textwidth]{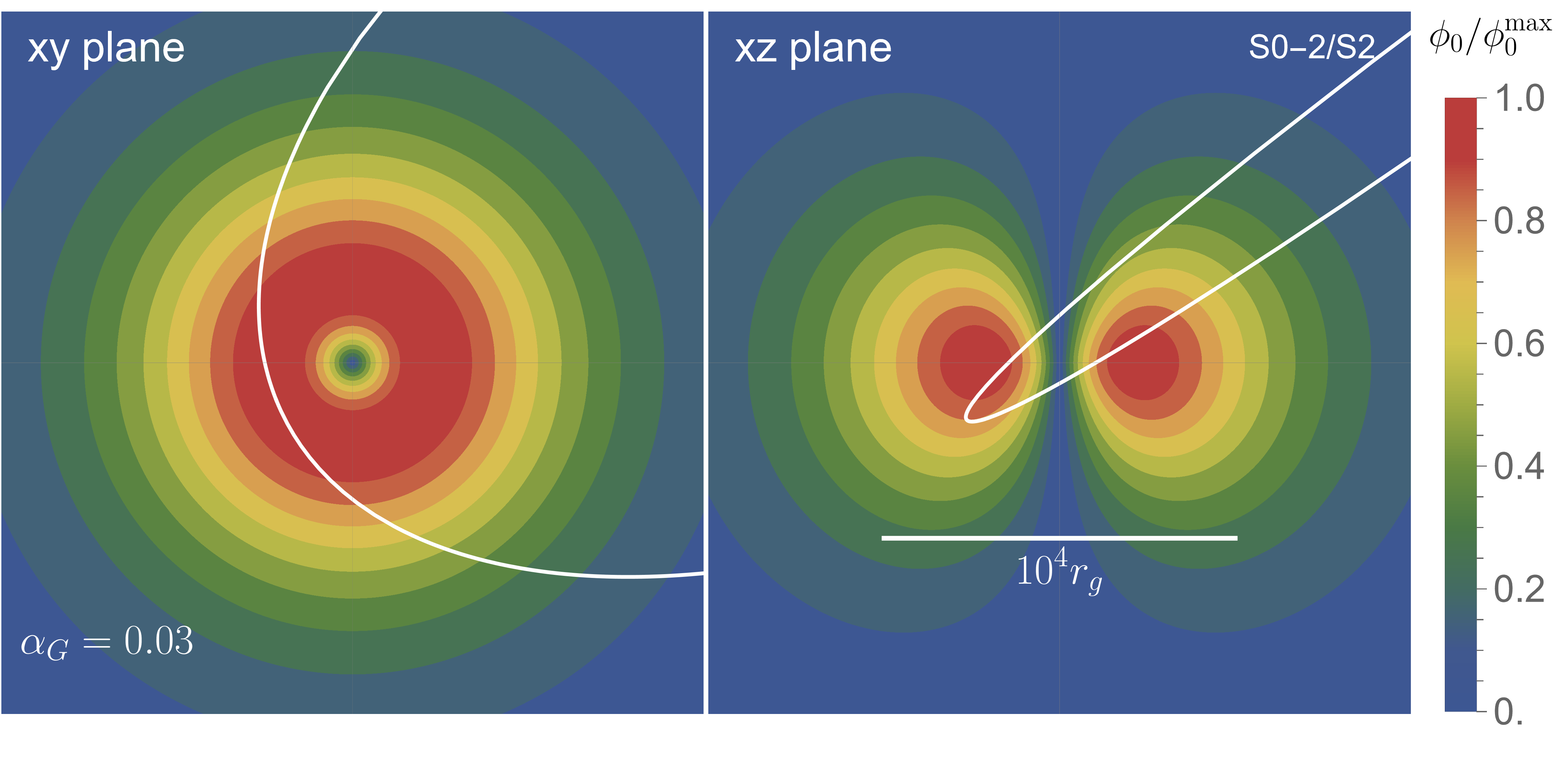}
\caption{An example of the normalized axion/scalar cloud density, $\phi_0/\phi_0^{\rm max}$, for $\alpha_G = 0.03$ on the $xy$-plane (left) and $xz$-plane (right) in the BH frame, where the BH spin axis is aligned with the $+z$ direction. The projected orbit of the S0-2/S2 star is also shown on both planes. The transformation of the S0-2/S2 orbit into the BH frame adopts fiducial values of $i_{\rm BH} = 155^\circ$ and $\Omega_{\rm BH} = 177^\circ$~\cite{GRAVITY:2023avo}.}
\label{fig:2DalphaEM}
\end{figure}

{Because the variation of the fine-structure constant depends on $\phi^2$, the signal contains a rapidly oscillating component at frequency $2\mu$. Substituting Eq.~(\ref{eq:phiwave}) and Eq.~(\ref{eq:phimax}) into Eq.~(\ref{eq:deltaalpha}) and using Eq.~(\ref{eq:phimax}) yields
\begin{equation}
\frac{\delta\alpha_{\rm EM}}
{\alpha_{\rm EM}}
=
\frac{1}{8} \alpha_G^2 C_\gamma
R(r)^2
\sin^2\theta
\left[
1+
\cos\!\left(
2\mu t-2\varphi+2\Delta_0
\right)
\right].
\label{eq:dalphaSRC}
\end{equation}
The first term corresponds to a static shift of the fine-structure constant, while the second term produces a coherent oscillation at frequency $2\mu$. In this work, we focus on the oscillatory component, which can be robustly distinguished from long-term astrophysical variability and orbital evolution.

Equation~(\ref{eq:dalphaSRC}) shows that the spatial dependence of the oscillation amplitude follows the squared cloud profile.
\begin{equation}
\left(
\frac{\phi_0}{\phi_0^{\rm max}}
\right)^2
=
R(r)^2\sin^2\theta.
\end{equation}
The corresponding profile is shown in Fig.~\ref{fig:2DalphaEM} on two orthogonal planes.

\subsection{Soliton Core Dark Matter}

Because the superradiant cloud wavefunction decays exponentially for
$r\gg1/(\mu\alpha_G)$, the corresponding modulation of the fine-structure
constant becomes strongly suppressed at large radii. An alternative scenario we consider is when ultralight scalars constitute dark matter and form a soliton-like core at the Galactic Center. For a typical virial velocity $v_{\rm vir}\sim10^{-3}$, the characteristic core radius is of order the de-Broglie wavelength~\cite{Schive:2014dra},
\begin{equation}
r_c
\simeq
4~{\rm pc}
\left(
\frac{10^{-18}{\rm eV}}{\mu}
\right).
\end{equation}

Inside the soliton core, the scalar field is approximately spatially uniform and can be written as
\begin{equation}
\phi
=
\phi_0^c
\cos(\mu t+\Delta_0),
\end{equation}
with amplitude
\begin{equation}
\phi_0^c
\simeq
2\times10^{11}\ {\rm GeV},
\label{eq:phi0core}
\end{equation}
which is largely independent of $\mu$ for $f_\phi\gg\phi_0^c$~\cite{Schive:2014dra,Yuan:2020xui}. The soliton may be regarded as a flattened inner Navarro-Frenk-White profile~\cite{Navarro:1995iw}, providing a conservative estimate of the central density, since additional mechanisms can further enhance the dark matter distribution near the Galactic Center~\cite{Kim:2022mdj,Budker:2023sex}.

Substituting the soliton solution into Eq.~(\ref{eq:deltaalpha}) yields
\begin{equation}
\frac{\delta\alpha_{\rm EM}}
{\alpha_{\rm EM}}
=
\frac{1}{2}
C_\gamma
\left(
\frac{\phi_0^c}{f_\phi}
\right)^2
\left[
1+
\cos\!\left(
2\mu t+2\Delta_0
\right)
\right].
\label{eq:dalphaCore}
\end{equation}
As in the superradiant case, the first term corresponds to a static shift of the fine-structure constant, while the second term produces a coherent oscillation at frequency $2\mu$.

The oscillatory component remains coherent throughout the soliton core. The coherence time,
$\sim1/(\mu v_{\rm vir}^2)$,
is much longer than the observation timescales considered in this work, ensuring a persistent signal shared by all stars residing within the core.}

\section{S-star Spectroscopic Measurements}

{The Galactic Center hosts a population of approximately 40 S-stars orbiting the SMBH Sgr A$^*$~\cite{Gillessen_2017}. Their atmospheres exhibit numerous atomic absorption lines whose frequencies depend sensitively on the fine-structure constant. Spectroscopic monitoring of these stars therefore provides a direct probe of temporal variations in $\alpha_{\rm EM}$~\cite{Hees:2020gda}.

For stars orbiting near an SMBH, the observed line frequencies are affected by both Doppler shifts and gravitational redshift. In the case of S-stars, whose orbital periods exceed 10 years, these effects evolve on timescales much longer than the scalar-induced oscillations considered here and can be accurately modeled and removed~\cite{Chu:2017gvt,Do:2019txf,Hees:2020gda}. Residual variations in the line positions therefore provide a direct probe of oscillations in the fine-structure constant.}

More explicitly, the residual deviation $\delta\lambda_j$ of the $j$-th spectral line relative to its rest wavelength $\lambda_j$ is related to variations in $\alpha_{\rm EM}$ through
\begin{equation}
\frac{\delta \lambda_j}{\lambda_j}
\simeq
-k_{\alpha,j}
\frac{\delta\alpha_{\rm EM}}
{\alpha_{\rm EM}},
\end{equation}
where $k_{\alpha,j}$ quantifies the sensitivity of the corresponding atomic transition to $\alpha_{\rm EM}$ and is tabulated in Ref.~\cite{Hees:2020gda}.

{We focus on the oscillatory component of $\delta\alpha_{\rm EM}$ with frequency $2\mu$. Owing to its periodic nature, this signal can be distinguished from the non-oscillatory component of the scalar field, which varies only on orbital timescales~\cite{Yuan:2022nmu}. The oscillation period is
\begin{equation}
T
=
\frac{2\pi}{2\mu}
\simeq
34.5~{\rm min}
\left(
\frac{10^{-18}\,{\rm eV}}
{\mu}
\right),
\end{equation}
which motivates high-cadence spectroscopic observations. The observational cadence limits the highest oscillation frequency that can be resolved. For example, a typical exposure time of $\sim10$ minutes implies sensitivity primarily to oscillation periods longer than about $20$ minutes.}

For our sensitivity analysis, we consider recent spectroscopic observations of S0-2/S2 obtained with the Gemini Near-Infrared Integral Field Spectrometer (NIFS) during its closest approach to the SMBH in 2017 and 2018~\cite{Do:2019txf}. The dataset consists of 25 observing epochs, each lasting approximately three hours.

The raw data comprise individual exposures with integration times of 10-15 minutes, which were subsequently averaged on a nightly basis in Ref.~\cite{Do:2019txf}. In our analysis, we instead treat the individual exposures separately in order to search for oscillatory signals on comparable timescales. To estimate the characteristic spectroscopic uncertainty per exposure, we analyze a representative night of NIFS observations, which naturally incorporates instrumental effects, telluric contamination, and stellar variability. Since the total observing duration is much longer than the expected oscillation period, stochastic astrophysical fluctuations can be efficiently distinguished from a coherent periodic signal. We then generate mock residual datasets for the hydrogen line over the 2017-2018 observing campaign, corresponding to the data reported in Ref.~\cite{Do:2019txf}, in the form of $\delta\lambda_j/\lambda_j$, assuming $k_{\alpha,j}\simeq2$ and a characteristic uncertainty of order $10^{-4}$ per exposure.

{Based on this setup, we derive projected constraints on both superradiant scalar clouds and soliton core dark matter by marginalizing over the unknown initial phase $\Delta_0$, as described in the following subsections and Supplemental Material.}

Future facilities such as the High-resolution Infrared Spectrograph for Exoplanet Characterization (HISPEC) at the Keck Observatory and the Multi-Objective Diffraction-limited High-Resolution Infrared Spectrograph (MODHIS) at the Thirty Meter Telescope are expected to substantially improve spectroscopic capabilities~\cite{Mawet:2019eed,10.1117/12.2681522}. These instruments are projected to achieve roughly three orders of magnitude better spectral precision together with higher observing cadence, with exposure times as short as five minutes. The improved cadence extends the accessible scalar-mass range, while the enhanced precision significantly increases sensitivity to oscillatory variations of the fine-structure constant.

In addition, these facilities will enable observations of fainter late-type stars closer to Sgr A$^*$. Compared to hot early-type stars such as S0-2/S2, late-type stars exhibit richer spectral features and lower intrinsic variability. Combining multiple spectral lines within a single star, as well as correlated measurements from multiple stars, therefore provides a promising route to suppress astrophysical systematics and further enhance sensitivity.

{\subsection{Constraints from a Superradiant Scalar Cloud}

We first consider a scalar superradiant cloud around Sgr A$^*$. For axions, the field amplitude saturates at the decay constant due to self-interactions, Eq.~(\ref{eq:phimax}), making $C_\gamma$ the primary parameter of interest. The accessible axion mass range is determined by the superradiance condition, Eq.~(\ref{eq:SRC}), together with the requirement that the cloud grows on astrophysically relevant timescales.

Since the superradiance condition is expressed in terms of the gravitational fine-structure constant $\alpha_G = G_N M_{\rm BH}\mu$, the BH mass is required to translate $\alpha_G$ into the corresponding axion mass. The mass of Sgr A$^*$ has been measured with high precision, $M_{\rm BH}=4.30\times10^6~M_\odot$, with an uncertainty of only $\pm0.25\%$ based on stellar-orbit measurements~\cite{GRAVITY:2021xju}.}

We adopt a conservative range
\be
0.025 \leq \alpha_G \leq 0.11,
\ee
for which the Newtonian approximation underlying Eqs.~(\ref{eq:phiwave}) and (\ref{eq:phimax}) remains valid. The superradiance condition, Eq.~(\ref{eq:SRC}), is satisfied throughout this range for $a_J\gtrsim0.5$, consistent with Event Horizon Telescope (EHT) indications that Sgr A$^*$ is rapidly rotating~\cite{EventHorizonTelescope:2022wkp,EventHorizonTelescope:2022urf}. The lower bound ensures that the superradiant growth timescale~\cite{Detweiler:1980uk},
\be
\tau_{\rm SR}\simeq \frac{24}{a_J\alpha_G^8\mu},
\ee
remains shorter than $\sim10^9$ years, compatible with both the estimated time since a possible major merger event involving Sgr A$^*$~\cite{Wang:2024lzu} and the age of the Universe~\cite{Planck:2018vyg}. The upper bound is determined by the observational cadence. This range corresponds to
\be
\mu=\frac{\alpha_G}{G_N M_{\rm BH}}
\in
[7.8\times10^{-19},\,3.4\times10^{-18}]~{\rm eV},
\ee
corresponding to modulation periods of the quadratic signal
\be
T\simeq\frac{\pi}{\mu}\in[10,44]~{\rm minutes}.
\ee
The extreme mass ratio between Sgr A$^*$ and the S-stars, together with their large separations $\gg10^3r_g$, ensures that the superradiant cloud is largely insensitive to tidal and environmental effects~\cite{Arvanitaki:2009fg,Baumann:2018vus,Cardoso:2020hca}. The very low accretion rate of Sgr A$^*$~\cite{EventHorizonTelescope:2022urf} further guarantees the stability of the BH mass and spin throughout the superradiant evolution.

Since the scalar cloud wavefunction, Eq.~(\ref{eq:phiwave}), is anisotropic and aligned with the BH spin axis, it is necessary to analyze S-star orbits in the BH reference frame. We follow the procedure of Ref.~\cite{Grould:2017bsw} to transform the orbital motion into the BH frame, as detailed in Supplemental Material. This transformation requires the BH inclination angle $i_{\rm BH}$ and spin-projection position angle $\Omega_{\rm BH}$. We adopt
\be
i_{\rm BH}=155^\circ\pm5^\circ,
\qquad
\Omega_{\rm BH}=177^\circ\pm25^\circ,
\ee
obtained from fits to hotspot motion~\cite{GRAVITY:2023avo}, consistent with previous GRAVITY and EHT measurements~\cite{2018,EventHorizonTelescope:2022wkp}. The orbit of S0-2/S2, one of the closest and best-characterized S-stars, is shown as the white curve in Fig.~\ref{fig:2DalphaEM}, based on the fiducial values.

{We first consider the Gemini/NIFS observations of S0-2/S2 during its pericenter passage in 2017 and 2018~\cite{Do:2019txf}. Since the cloud wavefunction decays exponentially for $r\gg1/(\mu\alpha_G)$, the strongest spectroscopic signals arise near periastron for highly eccentric orbits. The finite exposure time of approximately 10 minutes restricts the accessible parameter space to $\alpha_G\lesssim0.055$.}

The projected constraints on $|C_\gamma|$ are shown in Fig.~\ref{fig:constraints}, with details provided in Supplemental Material. The exclusion curve exhibits a shallow minimum below $|C_\gamma|=1$, reflecting the $\alpha_G^2$ suppression in Eq.~(\ref{eq:phimax}) at lower masses and the decreasing overlap with the peak cloud density at larger masses. The black curve corresponds to the central values of $i_{\rm BH}$ and $\Omega_{\rm BH}$, while the gray band indicates the associated uncertainties, which have only a minor impact on the overall sensitivity. For comparison, we also show the SN1987A constraint $\Lambda\equiv f_\phi/|C_\gamma|^{1/2}\gtrsim3~{\rm TeV}$~\cite{Olive:2007aj}, displayed on the right-hand axis and related to the left-hand axis for a fixed value of $f_\phi$.

\begin{figure}[h]
    \centering
    \includegraphics[width=0.75\textwidth]{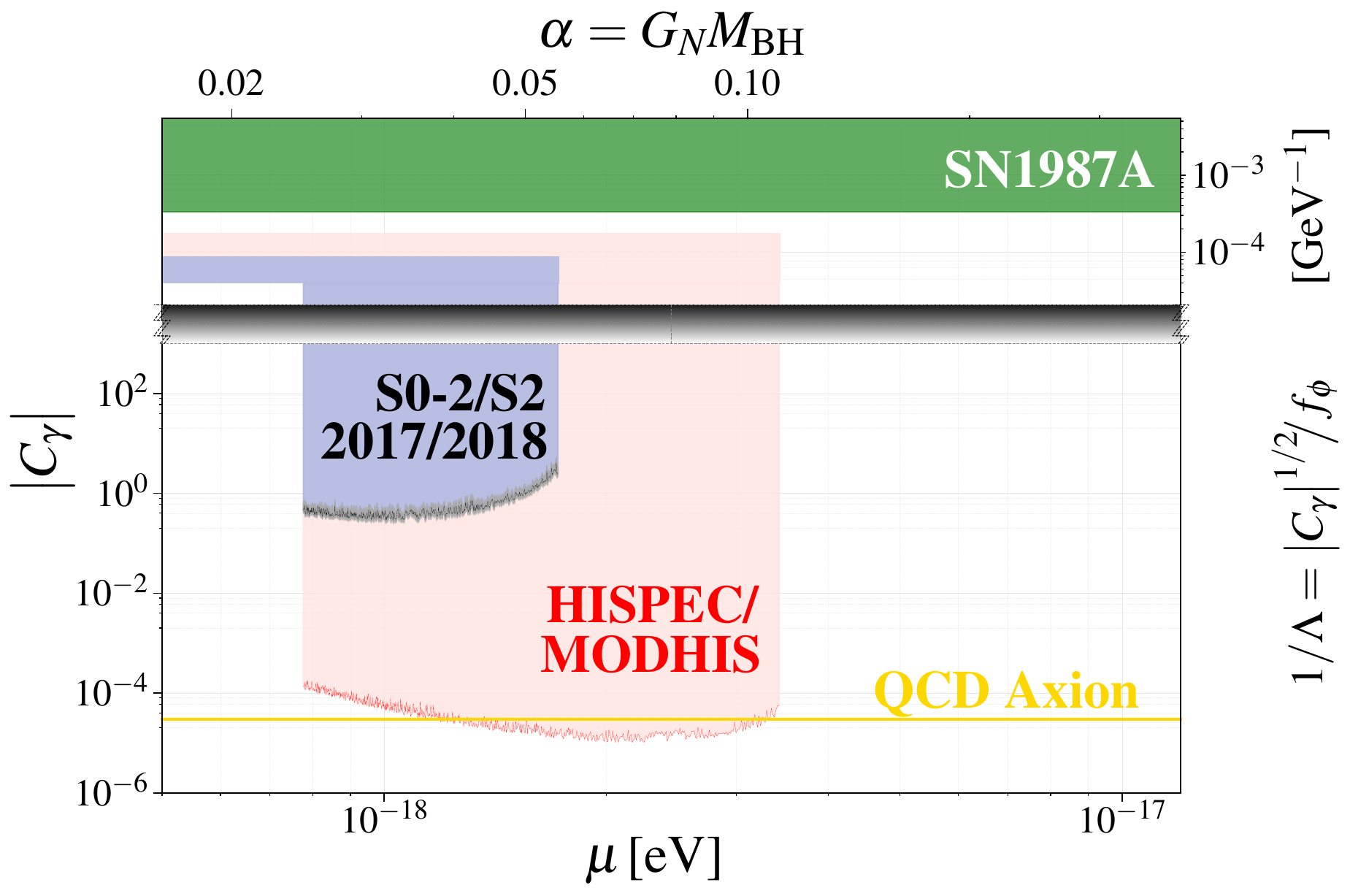}
    \caption{Projected constraints on the quadratic scalar–photon coupling $C_\gamma$ from recent spectroscopic observations of S0-2/S2 with Gemini/NIFS over 25 days in 2017–2018~\cite{Do:2019txf}, together with future prospects from HISPEC and MODHIS~\cite{Mawet:2019eed,10.1117/12.2681522}. The black line shows central values $i_{\rm BH}=155^\circ\pm5^\circ$ and $\Omega_{\rm BH}=177^\circ\pm25^\circ$~\cite{GRAVITY:2023avo}, with gray bands indicating their uncertainties. Future HISPEC/MODHIS projections assume a $10^3$ improvement in spectral resolution, halved cadence~\cite{Mawet:2019eed,10.1117/12.2681522}, a tenfold increase in observing time, and a late-type star with semi-major axis $0.1$ that of S0-2/S2. The range $0.025\le\alpha\le0.055\,(0.11)$ is set by the superradiant timescale and observational cadence. The horizontal line at $|C_\gamma|\simeq3\times10^{-5}$ corresponds to KSVZ-like QCD axions. Existing constraints from SN1987A~\cite{Olive:2007aj} are also shown, with the right axis corresponding to $\Lambda\equiv f_\phi/|C_\gamma|^{1/2}\ge3$~TeV. For $\Lambda\sim\mathrm{TeV}$, an mG-level background magnetic field induces an effective scalar mass and distorts the constrained mass window, as discussed in Supplemental Material.
    }
    \label{fig:constraints}
\end{figure}

In Fig.~\ref{fig:constraints}, we also present a future projection that includes a late-type star with one-tenth the semi-major axis of S0-2/S2 and a similar eccentricity of $0.8$, orbiting in the equatorial plane of the BH, with a total observation duration $10$ times longer than the data collected for S0-2/S2 in 2017 and 2018. The extension of the upper mass reach by a factor of two {$(\alpha \lesssim 0.11)$} is enabled by the improved observational cadence, while the enhanced sensitivity is driven primarily by the anticipated three-order-of-magnitude improvement in spectral resolution.

Reaching sensitivities of $|C_\gamma|\sim3\times10^{-5}$ would begin to probe the QCD axion parameter space for decay constants not already excluded by SMBH spin measurements, namely $f_\phi\lesssim10^{16}$~GeV, where such values can arise naturally in models with a $\mathbb{Z}_N$ symmetry~\cite{Hook:2018jle}. We note, however, that QCD axions in this mass range are already subject to strong constraints from axion-potential flipping in dense nucleon environments~\cite{Hook:2017psm,Zhang:2021mks,Balkin:2022qer,Gomez-Banon:2024oux,Gan:2025nlu}. Even for generic ALPs with quadratic photon couplings, higher-order effects can induce ALP–nucleon interactions that restrict the viable range of $f_\phi$. Such constraints may nevertheless be relaxed in the presence of additional sectors~\cite{Budnik:2020nwz,Madge:2024aot}.

{
We finally comment on the role of leading dimension-five axion couplings in superradiant clouds, compared with the quadratic photon coupling considered in this work. Although the dimension-six operator is formally suppressed by one additional power of the scalar field, this suppression is substantially reduced in a superradiant cloud, where the field amplitude can approach the symmetry-breaking scale. In the saturated regime, the quadratic signal carries only an additional factor of order $\phi_0/f_\phi\sim{\cal O}(\alpha_G)$ relative to a linear coupling, rather than an arbitrarily small perturbative suppression.

For the QCD axion, the dominant dimension-five interaction is the gluonic coupling, which induces an oscillating effective $\theta$ angle and hence time-dependent hadron and meson masses~\cite{Kim:2022ype}. In practice, however, these effects leave much weaker and less distinctive imprints on stellar spectral lines than oscillations of $\alpha_{\rm EM}$, and are considerably harder to isolate, since hadronic mass variations enter atomic spectra through multiple correlated quantities and are subject to significant degeneracies with stellar-atmosphere modeling, elemental abundances, and isotope-dependent effects.
Similarly, a linear axion-photon coupling would primarily induce polarization rotation ~\cite{Chen:2019fsq,Chen:2021lvo,Chen:2022oad} rather than the spectroscopic modulation considered here. Its coefficient is controlled by an independent anomaly coefficient and is therefore not necessarily related to the quadratic photon coupling. Moreover, the achievable sensitivity is limited by polarization systematics, including uncertainties associated with propagation effects, magnetic-field modeling, and the intrinsic degree of linear polarization.
For these reasons, the quadratic photon coupling provides a particularly clean and comparatively less degenerate observable.}

{\subsection{Constraints from Soliton Core Dark Matter}

We finally consider projected constraints on soliton core dark matter, as shown in Fig.~\ref{fig:axionDMconstraints}. We present limits on $1/\Lambda$ (right $y$-axis) for a generic quadratic scalar coupling and on $1/f_\phi$ (left $y$-axis) assuming the benchmark value $|C_\gamma|=|C_\gamma^{\rm QCD}|\simeq 3\times10^{-5}$ generated by an axion-gluon coupling.

For $\mu \lesssim 10^{-18}$~eV, the soliton core extends over parsec scales and encompasses the entire observed S-star cluster. Combined with the large field of view of future facilities, this enables simultaneous observations of approximately $40$ stars within a single coherent dark matter configuration. The coherent nature of the soliton core naturally facilitates a network of correlated spectroscopic measurements, in which multiple stars and multiple spectral lines can be analyzed jointly to search for a common oscillatory signal. {This coherence provides a unique advantage by allowing many spatially separated stellar probes to sample the same high-density dark matter configuration.}

For comparison, we also show existing terrestrial constraints on oscillations of the fine-structure constant, displayed as solid curves, including atomic-clock measurements~\cite{VanTilburg:2015oza,Zhang:2022ewz,Hees:2016gop,Kennedy:2020bac,Filzinger:2023zrs,Sherrill:2023zah} and fifth-force/equivalence-principle-violation (EPV) tests~\cite{Wagner:2012ui,Berge:2017ovy,Hees:2018fpg}. These EPV bounds have recently been refined to incorporate momentum-dependent effects arising from couplings to nucleons~\cite{Banerjee:2025dlo,Gue:2025nxq}. We also include a future projection from HISPEC/MODHIS, assuming a setup similar to that of Fig.~\ref{fig:constraints} but incorporating simultaneous observations of $40$ late-type stars. This configuration is expected to achieve comparable sensitivity at higher frequencies, providing complementary reach under conservative assumptions for the dark matter distribution.

Additional constraints arising from axion-gluon couplings are shown as dashed curves, including limits from SN1987A~\cite{Springmann:2024ret}, nuclear-parameter oscillations~\cite{Kim:2022ype,Banerjee:2023bjc,Madge:2024aot,Fuchs:2024xvc}, neutron electric dipole moment (nEDM) searches~\cite{Abel:2017rtm,Roussy:2020ily,Schulthess:2022pbp}, and nuclear EPV tests~\cite{Gue:2025nxq}. Additionally, cosmological constraints from Big Bang nucleosynthesis~\cite{Blum:2014vsa,Bouley:2022eer} and limits arising from nucleon environment backreaction~\cite{Hook:2017psm,Zhang:2021mks,Balkin:2022qer,Gomez-Banon:2024oux,Kumamoto:2024wjd} provide further restrictions on the parameter space. Figure~\ref{fig:axionDMconstraints} focuses on laboratory and astrophysical constraints, while relic-abundance considerations depend on the underlying cosmological history and are not shown.

\begin{figure}[t]
    \centering
    \includegraphics[width=0.75\textwidth]{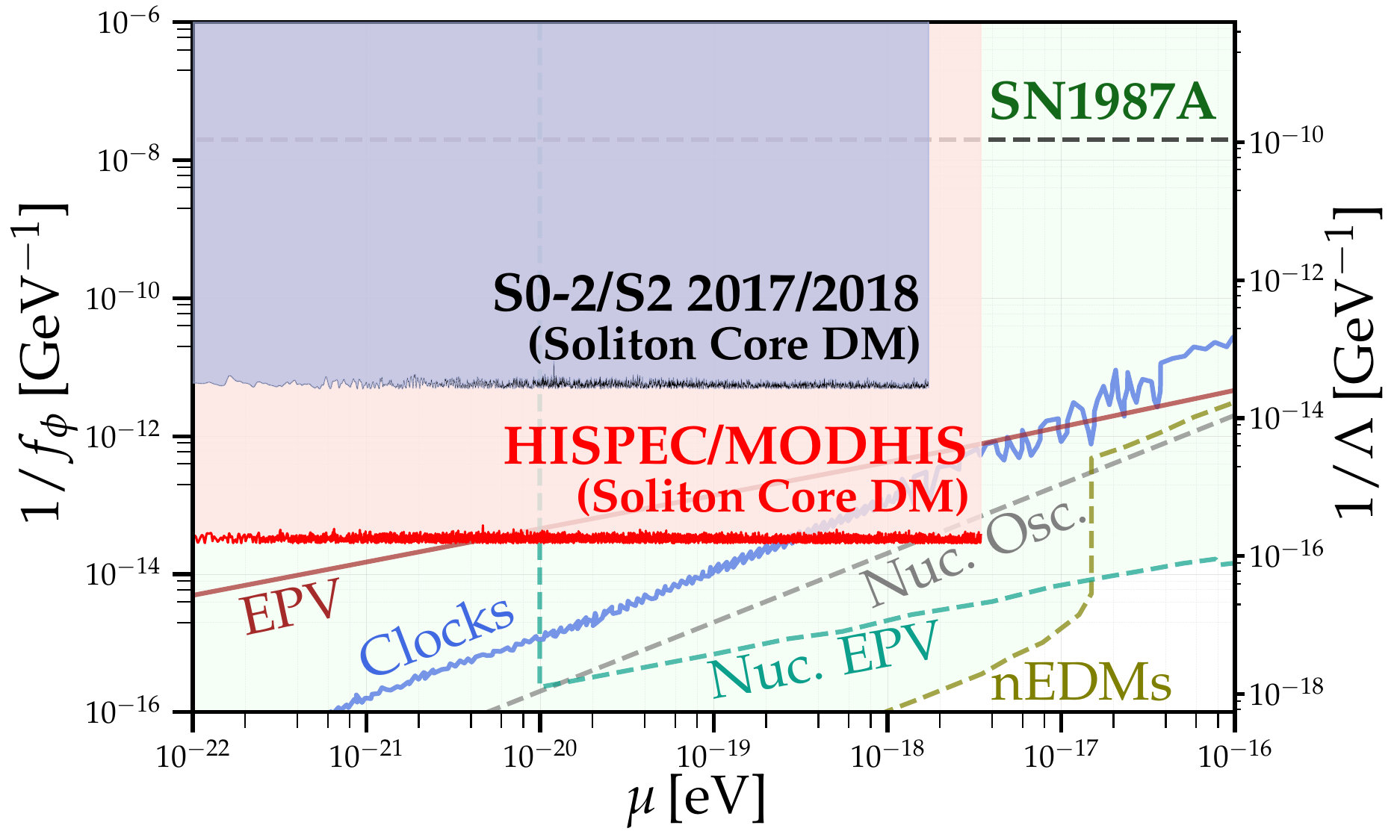}
    \caption{Projected constraints on the axion decay constant $f_\phi$ (left axis) assuming the benchmark loop-induced coupling $|C_\gamma|=|C_\gamma^{\rm QCD}|\simeq3\times10^{-5}$ generated by an axion–gluon interaction, and on the quadratic scalar–photon scale $\Lambda$ (right axis) for generic scalars, assuming DM with a core of nearly constant amplitude $\phi_0^c\simeq2\times10^{11}$~GeV. These constraints are obtained from the same S-star spectroscopic observations as in Fig.~\ref{fig:constraints}, with future HISPEC/MODHIS projections assuming simultaneous observations of $40$ late-type stars. Existing bounds from atomic-clock measurements of fine-structure constant variations~\cite{VanTilburg:2015oza,Zhang:2022ewz,Hees:2016gop,Kennedy:2020bac,Filzinger:2023zrs,Sherrill:2023zah} and from fifth-force/EPV tests~\cite{Wagner:2012ui,Berge:2017ovy,Hees:2018fpg} are shown as blue and brown lines. Additional limits from axion–gluon couplings, including SN1987A~\cite{Springmann:2024ret}, oscillating nEDMs~\cite{Abel:2017rtm,Roussy:2020ily,Schulthess:2022pbp}, nuclear parameter oscillations~\cite{Kim:2022ype,Banerjee:2023bjc,Madge:2024aot}, and nuclear EPV tests~\cite{Gue:2025nxq}, are indicated by dashed lines.
    }
    \label{fig:axionDMconstraints}
\end{figure}

\section{Discussion}
The environment near an SMBH at the Galactic Center can host extremely dense populations of ultralight bosons. Two representative scenarios are (i) BH superradiance, which exponentially amplifies a bound boson condensate and can drive the axion field to saturation near its decay constant, and (ii) ultralight DM forming a soliton core. The large field amplitudes in both cases substantially enhance axion-induced observables relative to terrestrial DM searches, making this system particularly sensitive to loop-induced interactions such as quadratic axion/scalar couplings to photons. In this work, we simulated precision spectroscopic measurements of S-stars orbiting Sgr A$^*$ to probe these couplings, considering axion profiles sourced by either a superradiant cloud or a soliton core. Current data already demonstrate sensitivity to sizable regions of parameter space, while near-future improvements in spectroscopic sensitivity promise substantially broader coverage. Beyond Sgr A$^*$, our framework applies to other BHs with companion stars or line-emitting plasma environments, enabling probes over many decades of scalar mass, including regimes relevant to vanilla QCD axions. Achieving this broader reach will require improved observational cadence, and the secular modulation of the bosonic cloud~\cite{Kim:2023pkx,Kim:2023kyy,Gan:2025icr} provides an additional probe of higher-mass regimes that complements sensitivity to rapid oscillatory signals.

While we focused on the Newtonian $(2,1,1)$ superradiant mode, other cloud profiles are also relevant. At larger $\alpha_G$, relativistic corrections become important and the $(3,2,2)$ mode may dominate the cloud mass and spatial extent, increasing stellar coverage~\cite{Omiya:2022gwu,Witte:2024drg}. Additional possibilities include spherical $(1,0,0)$ gravitational atoms formed from DM relaxation~\cite{Banerjee:2019epw,Budker:2023sex,Gan:2023swl}, as well as gravitational molecules corotating with comparable-mass binaries~\cite{Ikeda:2020xvt,Liu:2021llm,Bamber:2022pbs,Guo:2023lbv,Aurrekoetxea:2023jwk,Guo:2024iye,Guo:2025pea}. These systems span diverse mass ranges and may exhibit nonlinear phenomena such as bosenova~\cite{Eby:2016cnq,Levkov:2016rkk,Budker:2023sex,Gan:2023swl,Takahashi:2024fyq,Aurrekoetxea:2024cqd}. Spectroscopic observations offer a novel means of visualizing these dynamics, effectively enabling tomographic reconstruction of the evolving bosonic wavefunction.

A range of complementary observations can further elucidate both macroscopic cloud properties and microscopic axion couplings. Axions with linear photon couplings rotate the plane of linear polarization, allowing EHT measurements to probe the parameter $c\equiv g_{a\gamma}f_\phi$ once saturation is reached~\cite{Chen:2019fsq,Yuan:2020xui,Chen:2021lvo,Chen:2022oad,Ayzenberg:2023hfw,Wang:2023eip}. Combined polarimetry and spectroscopy thus provide a pathway to testing the ultraviolet origin of axions, alongside BH spin measurements that constrain large decay constants~\cite{Arvanitaki:2010sy,Arvanitaki:2014wva,Brito:2014wla,Cardoso:2018tly,Davoudiasl:2019nlo,Unal:2020jiy,Cheng:2022jsw,Saha:2022hcd,Guo:2024dqd,Guo:2025dkx}. Finally, purely gravitational observables can calibrate the cloud morphology through measurements of the cloud mass~\cite{Cunha:2019ikd,Chen:2022nbb,Chen:2022kzv,Shen:2023kkm,GRAVITY:2023cjt,Khalaf:2024nwc,GRAVITY:2024tth}.

\hspace{5mm}

\begin{acknowledgments}
We are grateful to Sebastian A. R. Ellis, Xucheng Gan, Stefan Gillessen, Yuxin Liu, Minyuan Jiang, Hyungjin Kim, Hidetoshi Omiya, Gilad Perez, Diogo Ribeiro, Matteo Sadun-Bordoni, Wolfram Ratzinger, Konstantin Springmann, Pham Ngoc Hoa Vuong, Sam Witte, Guan-Wen Yuan, Qiang Yuan, and Yue Zhao for useful discussions. 
The Center of Gravity is a Center of Excellence funded by the Danish National Research Foundation under grant No. 184. V.C. and Y.C. acknowledge support by VILLUM Foundation (grant no. VIL37766) and the DNRF Chair program (grant no. DNRF162) by the Danish National Research Foundation. V.C. is a Villum Investigator and a DNRF Chair.  V.C. acknowledges financial support provided under the European Union’s H2020 ERC Advanced Grant “Black holes: gravitational engines of discovery” grant agreement no. Gravitas–101052587.
Views and opinions expressed are however those of the author only and do not necessarily reflect those of the European Union or the European Research Council. Neither the European Union nor the granting authority can be held responsible for them. This project has received funding from the European Union's Horizon 2020 research and innovation programme under the Marie Sklodowska-Curie grant agreement No 101007855 and No 101131233. Y.C. is supported by the Rosenfeld foundation in the form of an Exchange Travel Grant and by the COST Action COSMIC WISPers CA21106, supported by COST (European Cooperation in Science and Technology).
IFAE is partially funded by the CERCA program of the Generalitat de Catalunya. X.X. is funded by the grant CNS2023-143767. 
Grant CNS2023-143767 funded by MICIU/AEI/10.13039/501100011033 and by European Union NextGenerationEU/PRTR.
Y.C. and X.X. acknowledge the support of the Rosenfeld foundation and the European Consortium for Astroparticle Theory in the form of an Exchange Travel Grant. H.X.~is supported by Fermi Forward Discovery Group, LLC under Contract No.\ 89243024CSC000002 with the U.S.\ Dept.\ of Energy, Office of Science, Office of High Energy Physics.
\end{acknowledgments}



\hspace{5mm}

\begin{center}
\textbf{\large Supplemental Material: Probing Fundamental Constant Oscillation in the Galactic Center with S-Star Spectroscopy}
\end{center}
\appendix
\setcounter{equation}{0}
\setcounter{figure}{0}
\setcounter{table}{0}
\numberwithin{equation}{section}
\counterwithin{figure}{section}
\counterwithin{table}{section}

\makeatletter
\@ifpackageloaded{hyperref}{%
  \renewcommand{\theHsection}{app.\Alph{section}}
  \renewcommand{\theHsubsection}{app.\Alph{section}.\arabic{subsection}}
  \renewcommand{\theHsubsubsection}{app.\Alph{section}.\arabic{subsection}.\arabic{subsubsection}}
  \renewcommand{\theHequation}{app.\Alph{section}.\arabic{equation}}
  \renewcommand{\theHfigure}{app.\Alph{section}.\arabic{figure}}
  \renewcommand{\theHtable}{app.\Alph{section}.\arabic{table}}
}{}
\renewcommand{\bibnumfmt}[1]{[#1]}
\renewcommand{\citenumfont}[1]{#1}
\makeatother

\section{S-Star Orbits in the Black Hole Frame}
This section provides details on how to specify the orbit of a star in the black hole (BH) frame, where the BH spin axis aligns with the $\hat{z}$-axis, following the procedures outlined in Ref.~\cite{Grould:2017bsw}.

\begin{figure}[h]
    \centering
    \includegraphics[width=0.75\textwidth]{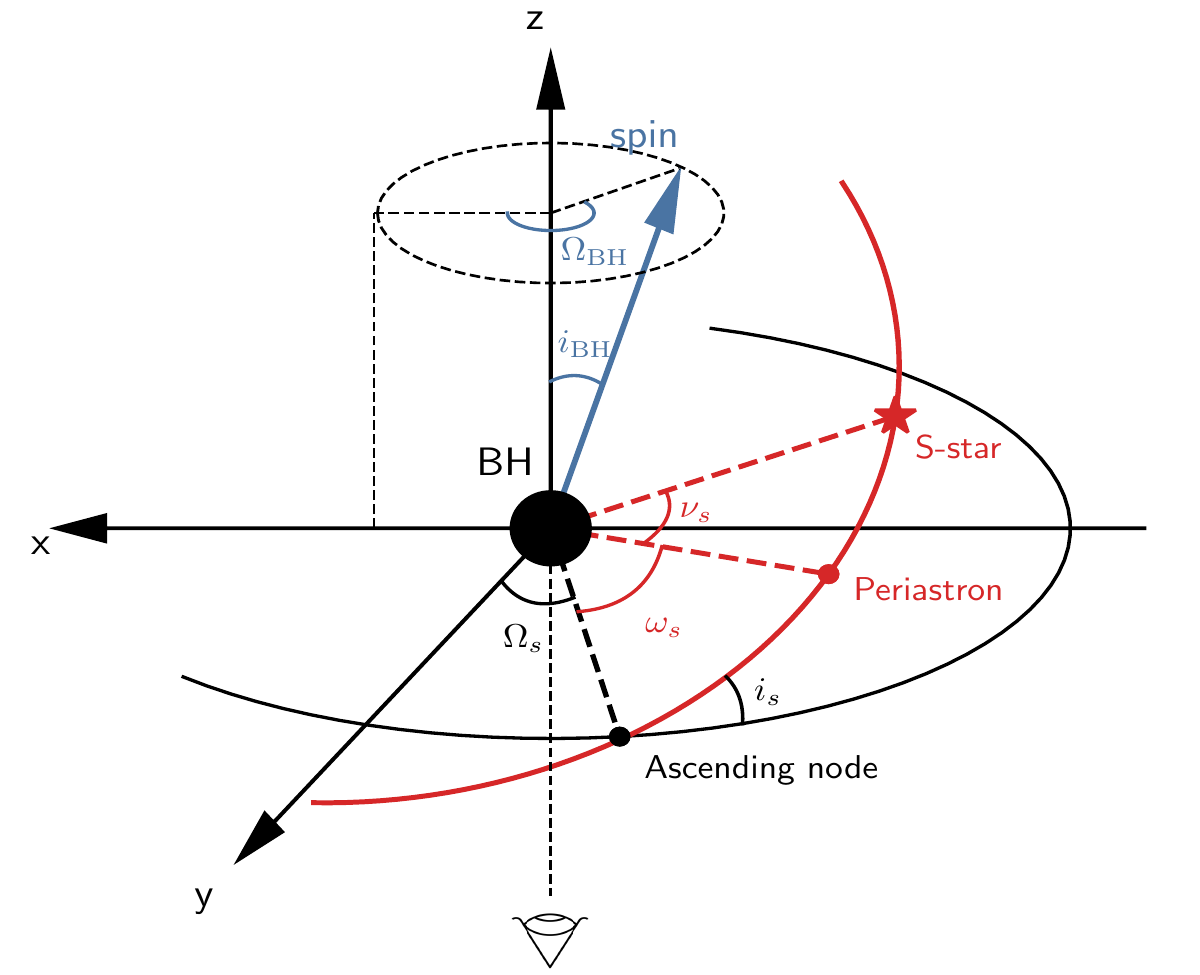}
    \caption{Illustration of the coordinate system in the observer's frame, where the $\hat{z}$-axis points from the observer to the orbit, and the $\hat{x}$-axis points to the north/declination. The star's orbit is depicted by the red line, while the BH spin direction is indicated by the blue arrow. The six orbital elements are defined in the discussion below, and the BH spin direction is determined by the inclination angle $i_{\rm BH}$ and the position angle $\Omega_{\rm BH}$.}
    \label{fig:BHCoor}
\end{figure}

The orbital motion is initially defined in the observer frame, where the $\hat{z}$-axis points from the observer to the orbit, and the $\hat{x}$-axis points to the north/declination, as illustrated in Fig.~\ref{fig:BHCoor}. The orbit is characterized by six orbital elements:
\begin{itemize}
    \item The eccentricity $e$ and semi-major axis $a$, which describe the size and shape of the orbit.
    \item The inclination $i_s$ and the longitude of the ascending node $\Omega_s$, which define the orientation of the orbital plane.
    \item The argument of periapsis $\omega_s$, which specifies the direction of the orbit's closest approach to the BH.
    \item The true anomaly $\nu_s$, which indicates the phase of the star along its orbit.
\end{itemize}
In the observer frame, the orbital motion can be expressed as
\begin{equation}
    \Vec{r}_s =  \frac{a(1 - e^2)}{1 + e \cos \nu_s} \left( \cos \nu_s \, \hat{a} + \sin \nu_s \, \hat{b} \right),
\end{equation}
where
\begin{equation}
    \hat{a} \equiv 
    \begin{pmatrix}
        -\cos \omega_s \sin \Omega_s - \sin \omega_s \cos \Omega_s \cos i_s \\
        \cos \omega_s \cos \Omega_s - \sin \omega_s \sin \Omega_s \cos i_s \\
        \sin \omega_s \sin i_s
    \end{pmatrix}, \quad
    \hat{b} \equiv 
    \begin{pmatrix}
        \sin \omega_s \sin \Omega_s - \cos \omega_s \cos \Omega_s \cos i_s \\
        -\sin \omega_s \cos \Omega_s - \cos \omega_s \sin \Omega_s \cos i_s \\
        \cos \omega_s \sin i_s
    \end{pmatrix}.
\end{equation}

To transform into the BH frame, we first consider the BH spin direction $\hat{J}_{\rm BH}$, represented by the blue arrow in Fig.~\ref{fig:BHCoor}. This direction is specified by the inclination angle $i_{\rm BH}$ and the position angle $\Omega_{\rm BH}$, expressed as
\begin{equation}
    \hat{J}_{\rm BH} = \left(\sin i_{\rm BH} \cos \Omega_{\rm BH}, \sin i_{\rm BH} \sin \Omega_{\rm BH}, \cos i_{\rm BH}\right).
\end{equation}
To align $\hat{J}_{\rm BH}$ with the $\hat{z}$-axis in the BH frame, we introduce a rotation matrix $\mathcal{R}_{\rm BH}$:
\begin{equation}
    \mathcal{R}_{\rm BH} = 
    \begin{pmatrix}
        \sin \Omega_{\rm BH} & -\cos \Omega_{\rm BH} & 0 \\
        \cos i_{\rm BH} \cos \Omega_{\rm BH} & \cos i_{\rm BH} \sin \Omega_{\rm BH} & -\sin i_{\rm BH} \\
        \sin i_{\rm BH} \cos \Omega_{\rm BH} & \sin i_{\rm BH} \sin \Omega_{\rm BH} & \cos i_{\rm BH}
    \end{pmatrix}.
\end{equation}

Using this transformation matrix, the star's orbit in the BH frame is given by
\begin{equation}
    \Vec{r}_{s,{\rm BH}} = \frac{a(1 - e^2)}{1 + e \cos \nu_s} \left( \cos \nu_s \, \hat{a}_{\rm BH} +  \sin \nu_s \, \hat{b}_{\rm BH}\right),
\end{equation}
where
\begin{equation}
    \hat{a}_{\rm BH} = \mathcal{R}_{\rm BH} \hat{a} = 
    \begin{pmatrix}
        -\cos \omega_s \cos (\Omega_{\rm BH} - \Omega_s) - \cos i_s \sin \omega_s \sin (\Omega_{\rm BH} - \Omega_s) \\
        \cos i_{\rm BH} \left[\cos \omega_s \sin (\Omega_{\rm BH} - \Omega_s)-\cos i_s \sin \omega_s \cos (\Omega_{\rm BH} - \Omega_s) \right] - \sin i_{\rm BH} \sin i_s \sin \omega_s \\
        \sin \omega_s \left[ \cos i_{\rm BH} \sin i_s-\sin i_{\rm BH} \cos i_s \cos (\Omega_{\rm BH} - \Omega_s) \right] + \sin i_{\rm BH} \cos \omega_s \sin (\Omega_{\rm BH} - \Omega_s)
    \end{pmatrix},
\end{equation}
and
\begin{equation}
    \hat{b}_{\rm BH} = \mathcal{R}_{\rm BH} \hat{b} = 
    \begin{pmatrix}
        \sin \omega_s \cos (\Omega_{\rm BH} - \Omega_s)-\cos i_s \cos \omega_s \sin (\Omega_{\rm BH} - \Omega_s)  \\
        -\cos i_{\rm BH} \left[\sin \omega_s \sin (\Omega_{\rm BH} - \Omega_s)+\cos i_s \cos \omega_s \cos (\Omega_{\rm BH} - \Omega_s) \right] - \sin i_{\rm BH} \cos i_s \sin \Omega_s \\
        -\sin i_{\rm BH} \left[\sin \omega_s \sin (\Omega_{\rm BH} - \Omega_s)+\cos i_s \cos \omega_s \cos (\Omega_{\rm BH} - \Omega_s) \right] + \cos i_{\rm BH} \cos i_s \sin \Omega_s
    \end{pmatrix}.
\end{equation}

\section{Statistics/Data Analysis}
\subsection{Data analysis}
The data consists of spectroscopic measurements indexed by $I$ (star), $n$ (night), and $i$ (exposure). The measured data $d_{I,n,i}$ includes contributions from a potential signal $s_{I,n,i}(\bm{\Theta})$ (dependent on parameters $\bm{\Theta}$) and Gaussian noise $n_{I,n,i} \sim \mathcal{N}(0, \sigma_{I,n,i}^2)$. The likelihood function is given by:
\begin{equation}
    \ln\mathcal{L}(\bm{\Theta}|\{d\}) = -\frac{1}{2}\sum_{I,n,i}\left(\frac{d_{I,n,i} - s_{I,n,i}(\bm{\Theta})}{\sigma_{I,n,i}}\right)^2 + \text{constant}.
\end{equation}
For a signal model linear in parameters ($s_a = \sum_\mu \mathcal{P}_{a,\mu} \Theta_\mu$, where $a \equiv \{I,n,i\}$), the maximum likelihood estimate is obtained by solving:
\begin{equation}
    \frac{\partial}{\partial \Theta_{\mu}}\ln\mathcal{L}(\bm{\Theta}|\{d\}) = 0,
\end{equation}
yielding:
\begin{equation}
    \begin{aligned}
    \bm{\Theta}_{\rm best-fit} = (\mathcal{P'}^T \mathcal{P'})^{-1} \mathcal{P'}^T \bm{b}, \quad \text{where } \ b_a \equiv \frac{d_a}{\sigma_a}, \quad \mathcal{P}'_{a,\mu} \equiv \frac{1}{\sigma_a}\mathcal{P}_{a,\mu}
    \end{aligned}
\end{equation}

Under the null hypothesis, mock data $d^{\rm mock}_a$ are generated as:
\begin{equation}
    d^{\rm mock}_{a} = n_{a}, \quad n_{a} \sim \mathcal{N}(0, \sigma_{a}^2).
\end{equation}
The marginalized posterior distribution of $\Theta_{\rm best-fit,\mu}^{\rm mock}$ derived from mock data follows:
\begin{equation}
    \Theta_{\rm best-fit,\mu}^{\rm mock} \sim \mathcal{N}\left(0, \sigma_{\Theta_\mu}^2\right), \quad \text{with } \sigma_{\Theta_\mu}^2 = \left[(\mathcal{P}'^T \mathcal{P}')^{-1}\right]_{\mu\mu},
\end{equation}
where $\sigma_{\Theta_\mu}$ is the theoretical uncertainty from the Fisher information matrix $\mathcal{I} = \mathcal{P}'^T \mathcal{P}'$.

An $x\%$ confidence level exclusion limit $\Theta_\mu^{\rm up}$ is defined as the value satisfying:
\begin{equation}
    P\left(\Theta_{\rm best-fit,\mu} \geq \Theta_\mu^{\rm up} \,|\, \Theta_\mu^{\rm true} = \Theta_\mu^{\rm up}\right) = (1-x)\%.
\end{equation}
For Gaussian-distributed $\Theta_{\rm best-fit,\mu}$, this corresponds to the $(1-x)\%$ quantile of the $\mathcal{N}(0, \sigma_{\Theta_\mu}^2)$ distribution. 

\subsection{Signal modeling}
From the main part of the paper, the spectroscopic signal due to ALP is
\begin{equation}
    \frac{\delta \lambda_j}{\lambda_j} = - k_{\alpha,j}\frac{\delta \alpha_{\rm EM}}{\alpha_{\rm EM}} = - k_{\alpha,j}C_\gamma \frac{\phi^2}{f_\phi^2} \, ,
\end{equation}
where the subscript $j$ refers to a particular atomic line, $k_{\alpha,j}$ is the sensitivity of the atomic transition to the fine structure constant (typically $\sim 2$ for H atomic lines).

Using the solution for the ALP scalar field $\phi$ provided in the main part of the paper, one finds that the signal can be written as
\begin{subequations}
    \begin{align}
        \frac{\delta \lambda}{\lambda} = s(t,\bm x) &= -k_{\alpha} C_\gamma \frac{\left(\sin \theta \phi_0^\mathrm{max} R(r)\right)^2}{2f_\phi^2}\left[1+\cos 2(\mu t-\varphi) \cos 2\Delta_0 -\sin 2(\mu t-\varphi) \sin 2\Delta_0 \right]\,  \\ 
        &= s_0(\bm x) \Big[A\sin(2\mu t - 2\varphi) + B\cos(2\mu t - 2\varphi)\Big] + \cdots\, ,
    \end{align}
    where $s_0(x) = k_\alpha \Big(\sin\theta \phi_0^{\mathrm{max}}R(r)\Big)^2/2$, and the $\cdots$ denotes terms evolving on the orbital timescale, which are not considered in this analysis focused on rapid oscillations. Note that in the data analysis, it is better to infer the $2$ parameters $A$ and $B$ since the model is linear in these parameters and to convert the resulting posterior into a posterior on the coupling parameter using $C_\gamma/f_\phi^2=\sqrt{A^2+B^2}$. 
    
\end{subequations}

\subsection{Example: Simplified Observational Scenario}
\label{subsec:example}
Here, for pedagogical purposes, we give an example of a data analysis in a simplified observational scenario. 

Consider a signal model of the form:
\begin{equation}
     s(t,\bm{x}) = s_0(\bm{x}) \Big[A\sin(2\mu t - 2\varphi) + B\cos(2\mu t - 2\varphi)\Big] + k_{I,n},
\end{equation}
where  $k_{I,n}$ represents a night-specific instrumental systematic offset for the night $n$. The free parameters are:
\begin{equation}
    \bm{\Theta} = \{A,\, B,\, k_{I,n}\}.
\end{equation}

We will consider a case where we observe 2 stars, with the following observation scheduling:
\begin{itemize}
    \item \textbf{Star 1}: 
    \begin{itemize}
        \item Night 1: 2 exposures (Obs a, Obs b)
        \item Night 2: 1 exposure (Obs c)
    \end{itemize}
    \item \textbf{Star 2}: 
    \begin{itemize}
        \item Night 1: 1 exposure (Obs d)
    \end{itemize}
\end{itemize}
This yields $4$ exposures and $5$ free parameters. The $\mathcal{P}'$-matrix (dimensions $4 \times 5$) becomes:
\begin{equation}
    \mathcal{P}' = \begin{pmatrix}
        \dfrac{s_0(\bm{x}_a)\sin\phi_a}{\sigma_a} & \dfrac{s_0(\bm{x}_a)\cos\phi_a}{\sigma_a} & \dfrac{1}{\sigma_a} & 0 & 0 \\
        \dfrac{s_0(\bm{x}_b)\sin\phi_b}{\sigma_b} & \dfrac{s_0(\bm{x}_b)\cos\phi_b}{\sigma_b} & \dfrac{1}{\sigma_b} & 0 & 0 \\
        \dfrac{s_0(\bm{x}_c)\sin\phi_c}{\sigma_c} & \dfrac{s_0(\bm{x}_c)\cos\phi_c}{\sigma_c} & 0 & \dfrac{1}{\sigma_c} & 0 \\
        \dfrac{s_0(\bm{x}_d)\sin\phi_d}{\sigma_d} & \dfrac{s_0(\bm{x}_d)\cos\phi_d}{\sigma_d} & 0 & 0 & \dfrac{1}{\sigma_d} \\
    \end{pmatrix},
\end{equation}
where $\phi_i \equiv 2\mu t_i - 2\varphi$, and subscripts $i=a,b,c,d$ denote observations (e.g., $\sigma_a = \sigma_{1,1,1}$, $\bm{x}_c = \bm{x}_{1,2,1}$).

\section{Generation of Mock data}
In this analysis, we explore two scenarios: (i) a case of existing data for the star S0-2/S2 and (ii) a case of futuristic measurements that will be performed using the next generation of instrument. 

\subsection{Existing data}
We consider the spectroscopic measurements of the short period star S0-2/S2 during the 2017-2018 measurement campaign, when the star was going through its periastron. The dataset is presented in Ref.~\cite{Do:2019txf} (see Sec.~$1.1$ from the Supplementary Materials of Ref.~\cite{Do:2019txf}). In 2017-2018, spectroscopic data has been taken from three instruments: (i) OSIRIS at the Keck observatory, (ii) IRCS at the SUBARU telescope and (iii)  NISF from the Gemini observatory. 

Each night of observation, several frames of integration time between 300 and 900\,s are taken. In Ref.~\cite{Do:2019txf}, all the frames for each night are combined together to produce a daily estimate of the star radial velocity (RV). In this analysis, since we are interested in measuring fast oscillations in the spectroscopic measurements such that it is more powerful to work directly with individual frames. This would require a detailed re-analysis of all raw data presented in Sec.~$1.1$ from the Supplementary Materials of Ref.~\cite{Do:2019txf}, which is beyond the scope of our sensitivity analysis. 

In order to perform our sensitivity analysis, we need to have a realistic estimate of the expected uncertainty of the spectroscopic measurement for each individual frame. In this sensitivity analysis, we use one night of data from the Gemini NIFS instrument and assume that the characteristic instrumental noise is representative of all spectroscopic measurements conducted between 2017 and 2019 across various instruments. Specifically, we consider the Gemini NIFS dataset from 2018-05-13, which is part of the data presented in Ref.~\cite{Do:2019txf}, consisting of 12 frames with 600\,s integration time each.
Of the 12 frames, only 8 of the individual data points were useful for individual relative RV measurements. The relative spectroscopic measurement is obtained  by cross-correlating each individual spectrum with a combined spectrum of the night. For simplicity, we only used the Hydrogen line at 2.1661\,microns (Br-$\gamma$). The individual extracted data is presented in Fig.~\ref{fig:S0-2_data}. The median RV uncertainty is $28$\,km/s and the RV scatter throughout the night is about $24$\,km/s, which is consistent with the uncertainties. The results can be sensitive to how the spectra are extracted and correlated so this could be refined in the future. 

\begin{figure}[t]
\centering
\includegraphics[width=0.75\linewidth]{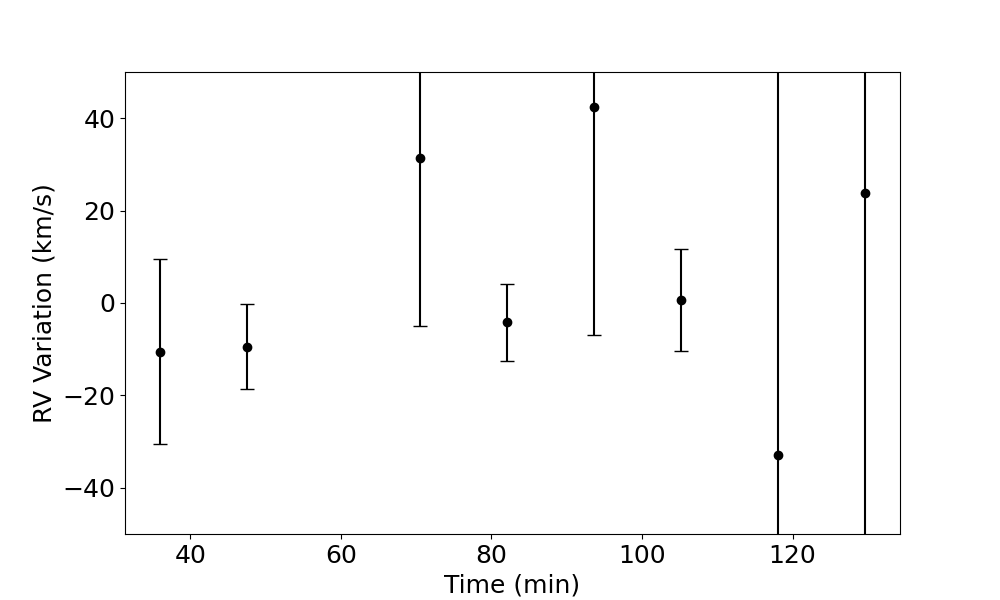}
\caption{Spectroscopic measurements of 8 out of the 12 frames of integration time of 600\,s taken on 2018-05-13 using the NIFS instrument on Gemini. The median RV uncertainty is 28\,km/s and the scatter throughout the nnight of 24\,km/s}
\label{fig:S0-2_data}
\end{figure}

In our analysis, we generated a mock dataset for 2017-2018 using the characteristic of the measurement noise from the night 2018-05-13. Specifically, we assume that 25\,\% of the frames will not be usable and we draw the RV uncertainty for each frame from a probability distribution that has been fitted to the distribution of the RV distribution from Fig.~\ref{fig:S0-2_data}; namely, $(\sigma_\mathrm{RV} - 10\,\mathrm{km/s}) / 10\,\mathrm{km/s}$ is modeled by a Gamma distribution with shape parameter $3$.

\subsection{Future data}
In our analysis, we also simulated data that would be representative of future instruments such as  HISPEC and MODHIS~\cite{Mawet:2019eed,10.1117/12.2681522}. With these instruments, we expect a three-order-of-magnitude improvement in spectroscopic resolution, a halved cadence~\cite{Mawet:2019eed,10.1117/12.2681522}. Such instruments would also enable the detection of late-type stars. With cooler atmospheres, these stars exhibit a rich variety of spectral lines with smaller intrinsic variance, in contrast to hot, early-type stars like S0-2/S2.
Thus, we assume a reduction of $\sigma_\mathrm{RV}$ by a factor of $10^3$.

We generate mock data incorporating these instrumental properties and spanning a time baseline ten times longer than the current dataset. To probe superradiant clouds, we consider a hypothetical late-type star with a semi-major axis equal to 10\% that of S0-2/S2, which would yield a significantly larger signal due to its closer proximity to the BH. For soliton-core dark matter, we assume simultaneous spectroscopic observations of $40$ stars within both the telescope’s field of view and the spatial extent of the soliton core.

\section{Plasma Effects}

A key question is whether the scalar cloud can continue to grow once corrections to the fine-structure constant become sizable. In vacuum, when the scalar field approaches the cutoff scale $\Lambda$, parametric annihilation into photons can efficiently deplete the cloud. In the astrophysical environment considered here, however, the photon plasma mass $\omega_p$ is several orders of magnitude larger than the scalar masses of interest, kinematically suppressing such annihilation channels. As a result, the dominant dissipation mechanism becomes plasma heating, analogous to constraints on low-mass dark photon dark matter~\cite{Dubovsky:2015cca}.

We consider the quadratic coupling $\phi^2 F_{\mu\nu}F^{\mu\nu}/(4\Lambda^2)$ and focus on the perturbative regime, where the induced variation of the fine-structure constant is small,
\begin{equation}
    \frac{\delta \alpha_{\rm EM}}{\alpha_{\rm EM}}
    \sim \frac{\phi_0^2}{\Lambda^2}
    \equiv \varepsilon \ll 1.
\end{equation}
Expanding around the background field $\phi=\phi_0+\delta\phi$, and considering a background magnetic field $\vec B_0$ and a transverse photon mode $A_T$ in a plasma, the linearized equations of motion in Fourier space $(\omega,\vec k)$ read
\begin{equation}
    \left[
    \vec k^{\,2} - \omega^2
    + \begin{pmatrix}
        \Omega_p^2 & \omega\,\dfrac{\phi_0}{\Lambda^2}B_0 \\
        \omega\,\dfrac{\phi_0}{\Lambda^2}B_0 & \mu^2
    \end{pmatrix}
    \right]
    \begin{pmatrix}
        A_T \\ \delta\phi
    \end{pmatrix}
    = 0 \,,
    \label{eq:eom-matrix}
\end{equation}
where $B_0\equiv|\vec B_0|$. The complex plasma frequency is
\begin{equation}
    \Omega_p^2 \equiv \frac{\omega_p^2}{1 + i\nu/\omega},
    \qquad
    \omega_p^2 = \frac{4\pi \alpha_{\rm EM} n_e}{m_e},
\end{equation}
with $n_e$ and $m_e$ the electron number density and mass. The electron–ion collision rate governing collisional damping is
\begin{equation}
    \nu
    = \frac{4\sqrt{2\pi}\,\alpha_{\rm EM}^2\,n_e}{3 m_e^{1/2} T_e^{3/2}}
      \log\Lambda_C
    \;\simeq\;
    2\times 10^{-23}~{\rm eV}
    \left(\frac{n_e}{10~{\rm cm}^{-3}}\right)
    \left(\frac{10^{7}~{\rm K}}{T_e}\right)^{3/2},
    \label{eq:nu-def}
\end{equation}
where we take the Coulomb logarithm $\log\Lambda_C \simeq 30$, and $T_e$ denotes the electron temperature.

Diagonalizing the $2\times2$ system in Eq.~\eqref{eq:eom-matrix} for small mixing and in the non-relativistic limit for the scalar ($|\vec k|\ll \omega \simeq \mu$), with $\mu \ll \omega_p$, one finds that the scalar eigenfrequency acquires an imaginary part inherited from the photon, corresponding to a damping rate
\begin{equation}
    \gamma_\phi^{\rm (plasma)}
    \;\simeq\;
    \frac{\phi_0^2}{\Lambda^4}\,
    \frac{B_0^2\,\nu}{2\omega_p^2}.
    \label{eq:gamma-phi-general}
\end{equation}

For the superradiant cloud considered in our work, the relevant region lies at radii
$r \gtrsim 100~r_g$, where the plasma is relatively dilute and the magnetic field corresponds to the large-scale Galactic Center environment rather than the near-horizon region.
As representative values we take~\cite{Crocker_2010,Crocker_2011,Hosseini_2020}
\begin{equation}
    n_e \sim 10~{\rm cm}^{-3}, 
    \qquad 
    T_e \sim 10^7~{\rm K}, 
    \qquad 
    B_0 \sim 1~{\rm mG}.
\end{equation}
These parameters yield $\nu \sim 2\times 10^{-23}~{\rm eV}$ and $\omega_p \sim 1\times 10^{-10}~{\rm eV}$.

Imposing the perturbative condition $\phi_0^2/\Lambda^2 = \varepsilon \ll 1$ and the current constraint $\Lambda \gtrsim 3~{\rm TeV}$, we obtain a conservative upper bound on the scalar cloud decay rate due to plasma heating,
\begin{equation}
    \gamma_\phi^{\rm (plasma)}
    \;\lesssim\;
    \frac{1}{\Lambda^2}
    \frac{B_0^2\,\nu}{2\omega_p^2}
    \;\sim\;
    3\times 10^{-38}~{\rm eV}.
\end{equation}

In our analysis, the mass windows of interest require the superradiant growth rate to exceed the inverse age of the Universe,
\begin{equation}
    \Gamma_{\rm SR} \gtrsim 10^{-32}~{\rm eV}.
\end{equation}
Therefore, even in the most conservative perturbative limit,
$\gamma_\phi^{\rm (plasma)} \ll \Gamma_{\rm SR}$, and collisional plasma heating cannot halt or significantly modify the exponential growth of the superradiant cloud. Consequently, the superradiant mass window remains unchanged.

We now consider the regime in which the scalar background continues to grow and the induced correction to the fine-structure constant becomes comparable to, or larger than, the vacuum contribution, while remaining below the fully nonperturbative QED limit $\delta\alpha_{\rm EM}\sim\mathcal{O}(1)$. For a superradiant cloud, this occurs only when $|C_\gamma|=f_\phi^2/\Lambda^2 \gg 1$. In this regime, the scalar evolution is still adiabatic compared to electromagnetic and plasma timescales, so the electromagnetic sector responds quasi-instantaneously to a slowly varying, scalar-dominated coupling.

Once $\alpha_{\rm EM}$ is no longer dominated by its vacuum value, plasma effects are significantly enhanced. In particular, the electron–ion collision rate and other dissipative processes increase with the effective electromagnetic coupling, while the photon plasma mass is also modified accordingly. As a consequence, plasma heating becomes increasingly efficient. This enhanced dissipation rapidly transfers energy from the scalar cloud into the plasma and leads to a saturation of the cloud configuration, in which the superradiant energy injection is balanced by plasma dissipation.

In this regime, the fine-structure constant exhibits adiabatic oscillations with amplitudes substantially larger than those allowed in vacuum. Such stable and large variations are already strongly constrained by the observed stability of atomic and molecular spectral lines in the Galactic Center.

For the scalar mass window, the region satisfying the superradiant condition and a sufficiently rapid growth rate remains the one constrained by our analysis, since the saturation phase extracts BH rotational energy linearly and induces only minor gravitational backreaction. Stronger coupling at smaller $\Lambda$ could marginally extend the low-mass end by reducing the threshold field value for saturation; for conservativeness, we do not include this possible extension.

On the other hand, a strong background magnetic field $B_0$ can modify superradiant dynamics by inducing an effective scalar mass through the quadratic interaction,
\begin{equation}
    \mu_{\rm eff}^2 \;=\; \mu^2 + \Delta \mu^2,
    \qquad
    \Delta \mu^2 \sim \frac{B_0^2}{\Lambda^2}
    \;\simeq\;
    \left(6 \times 10^{-18}\,\mathrm{eV}\right)^2
    \left(\frac{B_0}{1~\mathrm{mG}}\right)^2
    \left(\frac{3~\mathrm{TeV}}{\Lambda}\right)^2,
    \label{eq:mueff}
\end{equation}
up to order-unity geometric factors.

Superradiance is governed by the effective scalar mass $\mu_{\rm eff}$ rather than the bare mass $\mu$. Accordingly, the superradiant condition and the associated growth rate should be evaluated using $\mu_{\rm eff}$. If $\mu_{\rm eff}$ falls within the superradiant window probed in our analysis, then scalar masses below the nominal lower edge of the vacuum mass window can also be excluded, since the enhanced effective mass shortens the superradiant timescale and enables efficient cloud growth. Conversely, if the induced effective mass exceeds the upper boundary of the superradiant window, set by the superradiant condition and observational cadence limitations, no additional constraint can be obtained, leading instead to a cutoff of the exclusion region at large coupling.

As seen from Eq.~(\ref{eq:mueff}), for Galactic Center magnetic fields at the mG level and $\Lambda$ near the current bound of $3~\mathrm{TeV}$, the induced effective mass lies within the mass range relevant to this study. Consequently, its impact on the exclusion region in Fig.~2 introduces a cutoff near the upper edge of the constrained parameter space. The same applies to the soliton-core dark matter scenario, where only the uppermost region is modified.

\providecommand{\href}[2]{#2}\begingroup\raggedright\endgroup


\end{document}